\documentstyle[12pt]{article}
\begin{document}
\baselineskip=12pt
\pagestyle{empty}
\begin{flushright}
MUTP/94/04
\end{flushright}
\today
\bigskip
\bigskip
\baselineskip=24pt
\begin{center}
\begin{large}
{\bf Diffeomorphisms of Curved Manifolds}\\
\end{large}
\bigskip
\medskip
{\em J.S. Apps} \\
\medskip
Department of Theoretical Physics \\
The University, Manchester M13 9PL \\
\bigskip
\medskip
{\bf ABSTRACT }
\end{center}
We obtain an expression for the curvature of the Lie group
SDiff$\cal M$ and use it to derive Lukatskii's formula for the case where $\cal
M$ is locally Euclidean. We discuss qualitatively some previous findings for
SDiff$S^{2}$ in conjunction with our result.

\newpage
\pagestyle{plain}
\pagenumbering{arabic}
\vspace{1in}

\section {Preliminaries}
We consider the Lie group SDiff$\cal M$ of volume-preserving
diffeomorphisms of a $d$-dimensional closed Riemannian manifold $\cal M$.
Details are in Arnold \cite {arnold}. The Lie
algebra is the set of solenoidal vector fields on $\cal M$, endowed with the
metric
\begin {equation}
\label {metic}
\langle U,V \rangle =\int_{\cal M} d^{d}x \sqrt{g}\:\:U\cdot V
\end {equation}
If the metric and fields are invariant under the right-action of the
diffeomorphism group, then we require the covariant derivative on SDiff$\cal M$
to satisfy
\begin {equation}
\nabla_{U} \langle V,W \rangle =0
\end {equation}
so that, if the metric on SDiff$\cal M$ is Riemannian,
\begin {equation}
\label {gskew}
\langle \nabla_{U} V,W \rangle +\langle V,\nabla_{U}W \rangle =0
\end {equation}
We define $D_{U}$ to be the covariant derivative on $\cal M$ in the direction
of the field $U$. Since
\begin {equation}
(D_{U} V)\cdot W +V\cdot (D_{U} W) =D_{U} (V\cdot W)
=U^{\mu}\partial_{\mu} (V\cdot W)
\end {equation}
we have
\begin {equation}
\label {dskew}
\langle D_{U} V,W \rangle +\langle V,D_{U} W \rangle =\int_{\cal M} d^{d}x
\sqrt {g}\:\: U^{\mu} \partial_{\mu} (V\cdot W) =0
\end {equation}
where the latter equality follows from integration by parts, together with the
 sole\-noidal condition $\partial_{\mu} (\sqrt{g}\:\:U^{\mu}) =0$.
Note that (\ref{dskew}) extends to the case where $V$ and $W$ have a non-zero
divergence.

We make the orthogonal decomposition
\begin {equation}
D_{U}V =P_{U}V +Q_{U}V
\end {equation}
where $P_{U}V$ is the solenoidal projection, and, from (\ref {dskew}),
satisfies
\begin {equation}
\label {pskew}
\langle P_{U}V,W \rangle +\langle V,P_{U}W \rangle =0
\end {equation}
for $V$ and $W$ solenoidal. $P_{U}V$ is therefore a vector on SDiff$\cal M$
which satisfies the criterion (\ref {gskew}) for a covariant derivative of
right-invariant fields. Additionally, since the Lie bracket
$[U,V] =D_{U}V-D_{V}U$ of two solenoidal
fields $U$ and $V$ is itself solenoidal, we have
\begin {equation}
P_{U}V -P_{V}U =[U,V]
\end {equation}
so that $P$ is the torsionless connection on SDiff$\cal M$, i.e.
$P_{U}V=\nabla_{U}V$. The curvature of
SDiff$\cal M$ with respect to fields $U$, $V$, $W$ and $X$ is then defined:
\begin {equation}
T_{UVWX}=\langle (P_{[U,V]}W -[P_{U},P_{V}]W) ,X\rangle
\end {equation}
We write this as
\begin {displaymath}
\langle [D_{[U,V]}W -D_{U}(D_{V}W-Q_{V}W) +D_{V}(D_{U}W-Q_{U}W)] ,X\rangle
\end {displaymath}
since $X$ is orthogonal to the $Q$ projection of $D$. Using (\ref {dskew}),
this becomes
\begin {displaymath}
\langle (D_{[U,V]}W -[D_{U},D_{V}]W) ,X\rangle -\langle Q_{V}W, Q_{U}X\rangle
+\langle Q_{U}W,Q_{V}X\rangle
\end {displaymath}
The first term is simply
\begin {displaymath}
\int_{\cal M} d^{d}x \sqrt{g}\:\: U^{\mu} V^{\nu} W^{\rho} X^{\sigma}
R_{\mu \nu \rho \sigma}
\end {displaymath}
where $R_{\mu \nu \rho \sigma}$ are the components of the Riemann tensor on
some coordinate basis on $\cal M$.

\section {Locally Euclidean Manifolds}
If $\cal M$ is locally Euclidean, then $R_{\mu \nu \rho \sigma}=0$. As
$Q_{U}V=Q_{V}U$, we obtain Lukatskii's formula \cite {lukatskii} for the
(unnormalized) sectional curvature $T_{UVUV}$:
\begin {equation}
T_{UVUV}
=\frac{1}{4}[\langle \hat{\chi}(U,U),\hat{\chi}(V,V)
\rangle -\langle \hat{\chi}(U,V),\hat{\chi}(U,V) \rangle]
\end {equation}
where $\chi (U,V)=D_{U}V+D_{V}U$, and the hat denotes the projection orthogonal
to SDiff$\cal M$.
Note that we suspect a typing error in equation (3) of \cite {lukatskii}, which
we have corrected: the ``$+$'' should be replaced by a ``$-$''.

\section {The General Case}
In all cases we can express $Q_{U}$V as the gradient of a single-valued
scalar field (see Arnold, p 341):
\begin {equation}
Q_{U}V =\hbox{grad}\:\xi \Rightarrow \hbox{div}\:(D_{U}V)=\bigtriangleup
\xi
\end {equation}
Writing $D_{U}=U\cdot \hbox{grad}$ and integrating by parts, our result is
\begin {eqnarray}
\label {its it}
T_{UVUV}&=&\int_{\cal M} d^{d}x \sqrt{g}\:\: \{\:\:U^{\mu} V^{\nu} U^{\rho}
V^{\sigma} R_{\mu \nu \rho \sigma}\nonumber \\
 &+&\hbox {div}\:[(U\cdot \hbox{grad}) V]
\bigtriangleup ^{-1} \hbox{div}\:[(U\cdot \hbox{grad}) V]\nonumber \\
&-&\hbox {div}\:[(U\cdot \hbox{grad}) U]
\bigtriangleup ^{-1} \hbox{div}\:[(V\cdot \hbox{grad}) V]\:\:\}
\end {eqnarray}
The sectional curvature of $\cal M$ in the plane defined by the vectors $U$ and
$V$ is given by
\begin {equation}
K_{\cal M}(U,V) =\frac{ U^{\mu}V^{\nu}U^{\rho}V^{\sigma} R_{\mu \nu \rho
\sigma} }
{ U^{\mu}V^{\nu}U^{\rho}V^{\sigma} (g_{\mu \rho} g_{\nu \sigma} -g_{\mu \sigma}
g_{\nu \rho}) }
\end {equation}
so that the first term in (\ref {its it}) is
\begin {displaymath}
\int_{\cal M} d^{d}x \sqrt{g}\:\:[(U\cdot U)(V\cdot V) -(U\cdot V)^{2}]K_{\cal
M}(U,V)
\end {displaymath}
Since $(U\cdot U)(V\cdot V) -(U\cdot V)^{2} \geq 0$, the sign of this term
is definite if the sign of $K_{\cal M}(U,V)$ is definite.

In particular, for the case $d=2$, $K_{\cal M}(U,V)$ is independent of $U$ and
$V$, and is equal to $R/2$, where $R$ is the Ricci scalar.
 Also, a solenoidal field on $\cal M$ is the curl of an (in general
multi-valued) scalar potential :
\begin {equation}
U^{\mu}= \frac{\epsilon^{\mu \nu}}{\sqrt{g}} \partial_{\nu} \chi
\end {equation}
If the potentials for $U$ and $V$ are $\chi$ and $\psi$ respectively
then the first term in (\ref {its it}) becomes
\begin {displaymath}
\int_{\cal M} d^{2}x \sqrt{g}\:\: \frac{1}{2} R \{\chi ,\psi \}^{2}
\end {displaymath}
where $\{ \chi ,\psi \}$ is the generally covariant Poisson bracket $(\epsilon
^{\mu \nu}\partial_{\nu} \chi \partial_{\mu} \psi)/\sqrt{g}$.

\section {The 2-sphere}
As an example, we calculate the sectional curvature of SDiff$S^{2}$,
alternative procedures for which have already been established by Arakelyan and
Savvidy \cite {arakelyan} and Lukatskii \cite {lukspher}. The sphere is of unit
radius, so that $R=2$. We choose the diffeomorphism generators
$U=\hbox{curl}\:(\cos \theta )$ and
$V=\hbox{curl}\:[\sin \theta (e^{i\phi}+e^{-i\phi})]$. For this simple case,
all relevant quantities turn out to be proportional to spherical harmonics, so
calculation of the inverse Laplacian is easy.
\begin {eqnarray}
\hbox{div}\:[(U\cdot \hbox{grad})U]&=&1-3\cos^{2}\theta \nonumber \\
\hbox{div}\:[(V\cdot \hbox{grad})V] &=& 2(3\cos^{2}\theta -1) \nonumber \\
&\Rightarrow & \bigtriangleup^{-1}\hbox{div}\:[(V\cdot \hbox{grad})V]
=-\frac{1}{3}(3\cos^{2}\theta -1) \nonumber \\
\hbox{div}\:[(U\cdot \hbox{grad})V] &=& -3\sin \theta \cos \theta
(e^{i\phi}+e^{-i\phi}) \nonumber \\
&\Rightarrow & \bigtriangleup^{-1}\hbox{div}\:[(U\cdot \hbox{grad})V]
=\frac{1}{2}\sin \theta \cos \theta (e^{i\phi}+e^{-i\phi})\nonumber
\end {eqnarray}
We obtain
\begin {eqnarray}
\label {one}
& &\int_{S^{2}} d^{2}x \sqrt{g}\:\: \frac{1}{2} R \{\chi ,\psi \}^{2}
= \frac{16\pi}{3}\\
\label {two}
& &\int_{S^{2}} d^{2}x \sqrt{g}\:\: \hbox {div}\:[(U\cdot \hbox{grad}) V]
\bigtriangleup ^{-1} \hbox{div}\:[(U\cdot \hbox{grad}) V]
= -\frac{8\pi}{5}\\
\label {three}
& &\int_{S^{2}} d^{2}x \sqrt{g}\:\: \hbox {div}\:[(U\cdot \hbox{grad}) U]
\bigtriangleup ^{-1} \hbox{div}\:[(V\cdot \hbox{grad}) V]
= \frac{16\pi}{15}
\end {eqnarray}
As $\langle U,U \rangle =8\pi /3$, $\langle V,V \rangle =32\pi /3$,
 $\langle U,V \rangle =0$, the normalized sectional curvature is
\begin {equation}
K_{S}(U,V)=
\frac{ \frac{16\pi}{3} -\frac{8\pi}{5} -\frac{16\pi}{15} }
{ \frac {8\pi}{3} \times \frac{32\pi}{3} }=\frac{3}{32\pi}
\end {equation}
This agrees with the result found by Lukatskii \cite {lukspher} and by Dowker
and Mo-zheng \cite {dowker} for the same vectors. In this instance, our method
 is much more tedious than the alternative, but this may not be true for a more
general case, in which summations over spherical harmonics are involved.

\section {Discussion}
The theory of volume-preserving diffeomorphism groups may be applied to fluid
 dynamics, where
the diffeomorphism generators give the velocity field of an incompressible
fluid on $\cal M$ (see \cite {arnold}). Given two flows with initial velocity
fields $U$ and $V$, a negative value of $K_{S}(U,V)$ causes
the difference in the fluid configurations to diverge with time. In this way,
the ``stability'' of a flow can be predicted.

We may construct a very heuristic argument concerning the origin of the
individual terms in (\ref {its it}), with reference to the fluid dynamical
interpretation. Since the second and third terms
arise from taking the solenoidal projection of the covariant derivative,
they can be seen as contributions due to forces between fluid elements which
maintain the volume-preserving nature of the flow. The first term, on the other
hand, only arises when $\cal M$ is curved, and may be interpreted as an effect
on the flow stability due to the deviation of geodesics on $\cal M$. We
therefore write
\begin {equation}
\label {split}
K_{S}(U,V) =K_{C}(U,V) +K_{F}(U,V)
\end {equation}
where $K_{C}(U,V)$ is the ``curvature'' term.

In the case of the sphere, there was some surprise (see \cite {dowker}) that
the sectional curvature $K_{S}(U,V)$ for
$U=\hbox{curl}\:(\cos \theta )$ turned out to be positive, since for similar
flows on a torus, the value is negative \cite {arnold}. In the light of the
preceeding argument, it seems more appropriate to compare only
$K_{F}(U,V)$ with Arnold's results, since the torus is flat.

We restrict ourselves to the case
\begin {eqnarray*}
U &=& \hbox{curl}\:(\cos \theta) \\
V &=& \hbox{curl}\: [Y_{l}^{m}(\theta ,\phi ) +(-1)^{m}Y_{l}^{-m}(\theta,
\phi)]
\end {eqnarray*}
so that $U$ and $V$ are real. It is easy to show that
\begin {displaymath}
\{ \cos \theta ,Y_{l}^{m}(\theta ,\phi )\}=imY_{l}^{m}(\theta ,\phi )
\end {displaymath}
and then use this to derive
\begin {equation}
K_{C}(U,V)=\frac{3m^{2}}{8\pi l(l+1)}
\end {equation}
The formula in \cite {lukspher} and \cite {dowker}:
\begin {equation}
K_{S}(U,V)=\frac{3m^{2}}{8\pi [l(l+1)]^{2}}
\end {equation}
then gives us
\begin {equation}
K_{F}(U,V)=\frac{3m^{2}}{8\pi l(l+1)} \left[ \frac{1}{l(l+1)}-1 \right]
\end {equation}
so that $K_{F}$ is always negative. We suggest that the positive curvature of
the sphere ``stabilizes'' the flow.

In general, we hope that equation (\ref {its it}) will facilitate the
calculation of the curvature of SDiff$\cal M$ for other manifolds. This is
certainly true for the torus: the use of Lukatskii's formula saves a great deal
of work in deriving Arnold's expression for the curvature.

I would like to thank Stuart Dowker for helpful discussions.

\end {document}